\documentclass[aps,prb,twocolumn,showpacs]{revtex4-1}
\usepackage{graphics}
\usepackage{epsfig}
\usepackage{dcolumn}
\usepackage{amsmath}
\usepackage{color}

\newcommand{\beq}{\begin{equation}}
\newcommand{\eeq}{\end{equation}}
\newcommand{\bea}{\begin{eqnarray}}
\newcommand{\eea}{\end{eqnarray}}
\newcommand{\bwt}{\begin{widetext}}
\newcommand{\ewt}{\end{widetext}}
\newcommand{\m}{\mathbf}

\begin{document}
\title{
Linear magnetoresistance from Dirac-like fermions in graphite
 }
\date{\today}
\author{Hridis K. Pal$^{1,2}$ and Dmitrii L. Maslov$^2$}

\begin{abstract}

We show that
magnetoresistance 
of
 Bernal-stacked graphite (with the magnetic field ${\bf B}$ parallel to the $c$-axis and the current in the $ab$ plane
) scales linearly with the magnetic field over an interval of classically weak 
fields.
 The linearity is
related to the presence of extremely light, Dirac-like carriers near the $H$ ($H^{\prime}$)- points of the Brillouin zone. The Hall resistivity  in this interval also shows a non-analytic, 
$B\ln |B|$
 behavior,  and is 
  dominated by holes. 
\end{abstract}

\affiliation{$^1$School of Physics, Georgia Institute of Technology, Atlanta, GA 30332-0430, USA\\
$^2$Department
of Physics, University of Florida,
Gainesville, FL 32611-8440, USA}

\maketitle
\section{Introduction}

Dirac fermions have become a focus of great interest in condensed matter physics in recent years. 
Being massless,
Dirac fermions
respond to the magnetic field much more strongly than their
massive counterparts. 
A popular platform for studying Dirac fermions is graphene, where their sensitivity to the magnetic field allows for the observation of the quantum Hall effect even at room temperature.\cite{geim}
Stacking up 
graphene layers in 
a Bernal (ABAB...) way,  one 
forms a three-dimensional crystal of graphite.
 For the most part of the Brillouin zone (BZ) and at energies smaller than interlayer hopping $(\sim 0.4$\;eV), the charge carriers in graphite have little resemblance
to their Dirac \lq\lq ancestors\rq\rq\/: they are just massive 
electrons and holes.
Graphenic
ancestry, however, 
makes 
the effective mass depends on the momentum along the $c$ axis
in such a way that it vanishes at the top and bottom edges of the BZ ($H (H')$- points). Tiny regions around these points harbor Dirac-like fermions 
similar to those 
in a single-layer graphene. Landau levels of these Dirac fermions in graphite have been observed by tunneling \cite{li} and optical \cite{chuang} spectroscopies. Their contribution to transport, however, is not clear;
in particular,
their role in quantum magnetoscillations has been a subject of recent discussion and remains controversial. \cite{kop2,pot} 

Although transport in graphite has been well studied in the past, there remain quite a few open questions. For example,  neither 
the 
magnitude nor the temperature or magnetic-field dependences of the $c$-axis conductivity can be satisfactorily explained within the Boltzmann transport theory.\cite{graphite,maslov,lmr,garcia,casparis} In comparison, the in-plane conductivity is relatively better understood; 
for example,  its
 temperature
dependence 
is well
explained using the semiclassical Boltzmann equation. \cite{ono,kim,inplane,kab1} The success, however, is only partial as the explanation of the field-dependence of in-plane magnetoresistance (MR) presents some difficulty. Experimentally,
in-plane magnetoresistance (with the magnetic field along the $c$-axis
and the electrical current in the $ab$-plane) 
is often found to depend linearly on the  magnetic field\cite{kab2,kop1,artunpub}
which, at first glance, seems to contradict the transport theory.
Even more surprisingly, linear MR  spans over a wide range of fields,  
beginning at classically weak fields and persisting up to the ultra-quantum regime and beyond.
\cite{graphite} 
Although the linearity in the ultra-quantum regime can be explained by 
taking into account 
the field dependence 
of  the scattering time,\cite{graphite} the linear behavior in semi-classical fields still lacks a proper
understanding. 
Although it is sometimes ascribed to 
extrinsic reasons, such as macroscopic inhomogeneities, 
the issue is far from being settled.\cite{kop1}

Since the magnetic field is a (pseudo) vector, MR can only be a function of $B^2$. In general,
MR behaves quadratically for weak fields and
either saturates,
if electron orbits are closed, or grows quadratically in strong fields,\cite{zim,abr} 
if orbits are open or 
in compensated metals.
 A linear and thus non-analytic dependence on the field 
indicates some non-trivial physics.
Experimentally, graphite is not the only material that exhibits linear MR---it is found to occur in many other
materials as well;\cite{linmagresexam} 
yet detailed understanding of this effect exists only in a handful of situations.\cite{linmagnature} For example, it has been shown that a non-analytic dependence on the field
can arise due to special features of the Fermi surface,\cite{leb,sch} which makes one wonder whether graphite 
has any such peculiarities too. 
In fact, previous studies of MR within the Boltzmann transport theory
 have already hinted 
 at such a possibility: numerical calculations have shown that at low temperatures and 
in weak fields, MR behaves as $
B^n$ with $n<2$, which indicates  a departure from the canonical behavior.\cite{ono,kab1}

In this paper, we show, using a simplified yet consistent model for the energy spectrum of charge carriers and semiclassical Boltzmann equation,
that linear MR is an inherent property of the graphite bandstructure.
The linearity 
stems
 from very light, Dirac-like carriers
near the $H (H')$- points of the BZ. 
The origin of the effect can be understood without detailed calculations. Indeed,
anywhere but in the immediate vicinities of the $H (H')$- points, electron-and hole-like carriers in graphite move in the $ab$-plane as free particles but with a mass that depends on the momentum along the $c$-axis ($k_z$). The corresponding spectrum is described by $\varepsilon_{\bf k}=f(k_z)\pm k_{\rho}^2/2m^*(k_z)$, where ${\bf k}_\rho$ is the in-plane momentum, $f(k_z$) is some function of $k_z$ and $m^*(k_z)\propto \cos(k_z c/2)$ with $c$ being the $c$-axis lattice constant.  The in-plane magnetoconductivity contains the usual factor $\left[1+\omega^2_c(k_z)\tau^2\right]^{-1}$ averaged over $k_z$ (here $\omega_c(k_z)=eB/m^*(k_z)$ and $\tau$ is the relaxation time).  Near the $H (H')$- points ($k_z=\pm \pi/c$), the effective mass vanishes as $\pi/c-k_z$, the cyclotron frequency diverges, and the integral
over $k_z$
 behaves as $|B|$ instead of 
 $B^2$. Likewise, the Hall conductivity contains an average of $\omega_c(k_z)\left[1+\omega^2_c(k_z)\tau^2\right]^{-1}$, which also diverges, albeit only logarithmically, near the $H (H')$- points. As a result, the Hall conductivity behaves as $B\ln|B|$. 
 
In materials with bandstructures simpler than that of graphite, the only relevant scale for classical MR is a characteristic magnetic field, $B_c$, beyond which the period of cyclotron motion becomes shorter than the relaxation time. In case of graphite, this field can be defined by the condition $eB_c\tau/m^*(0)=1$. However, nonanalytic MR sets in at a much weaker field, $B_0$. To estimate $B_0$, we notice that the width of the region near the $H$ ($H'$) point that contributes to non-analytic MR is determined by the condition $eB_c\tau/m^*(k_z)\sim 1$. The effective mass at distance $\delta k_z$ from the $H$ ($H'$) point  is small in proportion to that distance: $m^*(k_z)\sim m^*(0)\delta k_z c$. The quadratic approximation of the energy spectrum breaks down when the energy of the in-plane motion, which is on the order of the Fermi energy,
becomes comparable to the nearest-plane hopping energy, $\gamma_1\cos(k_zc/2)\sim \gamma_1\delta k_z c$. Estimating $\delta k_z$ from the last condition, we obtain $B_0\sim (\varepsilon_F/\gamma_1)B_c$. A special property of graphite, which makes it a low-density semimetal, is that $\varepsilon_F$ is fixed by the {\em next-to-nearest} inter-plane hopping, which is significantly smaller than $\gamma_1$ (see Sec.~\ref{sec2}).  Within the conventional bandstructure model,  we find $B_0\approx 0.06 B_c$.

Using the full spectrum instead of the quadratic approximation, one can show that MR is analytic for $B\ll B_0$. In classically strong fields ($B\gg B_c$), MR is again analytic.
Thus, non-analytic MR occurs
 in intermediate field range, i.e., for 
 $B_0\ll B\ll B_c$, and the interval of fields between $B_c$ and $B_0$ is sufficiently wide.
In experiment, however, linear MR often spans the entire interval of magnetic field: from the weak-field regime to the ultra-quantum limit.
Therefore, intrinsic linear MR, predicted in this paper, cannot explain the data for all magnetic fields. It is possible that other factors, such as macroscopic inhomogeneities, are responsible for linear MR in strong magnetic field. 

The rest of the paper is organized as follows. Sec.~\ref{sec2} describes the model for the electronic spectrum of graphite
which allows for an analytic calculation of the conductivity.
In Sec.~\ref{sec3}, we calculate the field dependences of the magnetoconductivity, first in the quadratic--\lq\lq non-relativistic\rq\rq\/ approximation (Sec.~\ref{subsec3.1}), and then for the full spectrum (Sec.~\ref{subsec3.2}).
The effect of macroscopic inhomogeneities is discussed in Sec.~\ref{subsec3.3.1}. 
An issue whether 
Dirac-like carriers in graphite play 
a role in quantum oscillations is addressed in Sec.~\ref{subsec3.3.2}. Our conclusions are given in Sec.~\ref{sec4}.

\section{\label{sec2} Energy spectrum of graphite}

Ideal graphite 
consists
of graphene planes stacked on top of each other in the Bernal way (ABAB...).
Its band structure is usually described by the Slonczewski-Weiss-McClure
(SWMc) model \cite{graphite} characterized by seven parameters: $\gamma_0\dots\gamma_
5$ and $\Delta$. Here, $\gamma_0$ and $\gamma_1$ denote the in-plane and out-of-plane nearest-neighbor hopping terms, correspondingly; $\gamma_2\dots\gamma_5$ denote various
next-nearest-neighbor hopping terms; and $\Delta$ arises due to the difference between the on-site energies of the A and B carbon atoms. In terms of energy scales,\cite{graphite} $\gamma_0\approx 3.2$ eV is the
largest one, followed by $\gamma_1\approx 0.4$ eV, $\gamma_3\approx 0.3$ eV, and $\gamma_4\approx 0.1$ eV, followed by $\gamma_2\approx -0.02$ eV, $\gamma_5\approx 0.01$ eV, and $\Delta\approx 0.01$ eV. A closed form of the
energy spectrum can be obtained only if $\gamma_3$ is put to zero. Under this approximation (discussed in more detail below), the energy spectra of the conduction and the valence bands, respectively, can be written as
\bea
\label{eq:spectrum}
\varepsilon_{\mathbf{k}}^{+}&=&\frac{1}{2}\left(\varepsilon_2^0+\varepsilon_3^0\right)+\left\{\frac{1}{4}\left(\varepsilon_2^0-\varepsilon_3^0\right)^2+v_{\rho}^2k_{\rho}^2\left(1+\frac{\gamma_4}{\gamma_0}\Gamma\right)^2\right\}^{\frac{1}{2}},\nonumber\\
\varepsilon_{\mathbf{k}}^{-}&=&\frac{1}{2}\left(\varepsilon_1^0+\varepsilon_3^0\right)-\left\{\frac{1}{4}\left(\varepsilon_1^0-\varepsilon_3^0\right)^2+v_{\rho}^2k_{\rho}^2\left(1-\frac{\gamma_4}{\gamma_0}\Gamma\right)^2\right\}^{\frac{1}{2}},\nonumber\\
\Gamma&=&2\cos(k_zc/2),
\eea
where 
$\varepsilon_{1,2}^0=\Delta
\pm
\gamma_1\Gamma+\frac{1}{2}\gamma_5\Gamma^2$,
$\varepsilon_3^0=\frac{1}{2}\gamma_2\Gamma^2$, 
$v_{\rho}=\frac{\sqrt{3}}{2}\gamma_0a$, and $a (c)$ is the in- (out-of-)
plane lattice constant.

\begin{figure}[t]
\includegraphics[angle=0,width=0.3\textwidth]{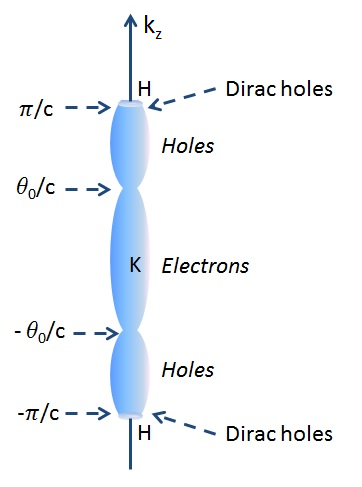}
\caption{Schematic representation of the Fermi surface of graphite.  Non-relativistic massive fermions occupy most of the Fermi surface, except for the narrow regions near the $H (H')$- points, occupied by massless Dirac fermions.
Here $\theta_0=\mathrm{cos}^{-1}\sqrt{\varepsilon_F/2\gamma_2}\approx 0.62$ 
marks the point at which the Fermi energy intersects the electron and hole bands.}
\label{fig_fermi}
\end{figure}

To facilitate an analytical calculation of the conductivity later, we 
simplify these expressions as much as possible, while keeping the
physical content intact. First, we note that the parameter $\gamma_4$, which enters as a ratio $\gamma_4/\gamma_0\approx 0.1$,  only introduces a weak $k_z$ dispersion 
in the in-plane velocity and thus can be neglected. Next, although the  parameters $\gamma_2$, $\gamma_5$, and $\gamma_6$ are of the same order of magnitude, their roles in the spectrum are very different.
 Namely, $\gamma_2$ plays a crucial role as it determines the band overlap, and 
hence small ($\approx 3\times 10^{18}$ cm$^{-3}$ as $T\rightarrow 0$) but non-zero
carrier concentration.
On the other hand, 
$\gamma_5$ and $\Delta$ do not lead to qualitative changes in the spectrum and thus can be neglected. With these simplifications, the minimal model for the energy spectrum in graphite can be written as
\bea
\label{eq:spectrum1}
\varepsilon_{\mathbf{k}}^{\pm}&=&\frac{1}{2}\left(\mp\gamma_1\Gamma+\frac{1}{2}\gamma_2\Gamma^2\right)\notag\\
&&\pm
\left\{\frac{1}{4}\left(\gamma_1\Gamma\pm\frac{1}{2}\gamma_2\Gamma^2\right)^2+v_{\rho}^2k_{\rho}^2\right\}^{\frac{1}{2}}.
\eea
(For $\gamma_2=0$, Eq.~(\ref{eq:spectrum1}) reduces to the original Wallace model\cite{Wallace} which describes graphite as a zero-gap semiconductor.) We will refer to the spectrum 
in Eq.~(\ref{eq:spectrum1}) as 
to
\lq\lq relativistic\rq\rq\/.
Equation (\ref{eq:spectrum1}) describes two groups of {\em massive} Dirac fermions with the \lq\lq rest mass\rq\rq\/
which vanishes at the $H (H')$- points $k_z=\pm \pi/c$, where $\Gamma=0$ (cf. Fig. \ref{fig_fermi}). Away from the $H (H')$- points one can expand Eq.~(\ref{eq:spectrum}) in $k_\rho$, which gives a \lq\lq non-relativistic\rq\rq\/ spectrum with the effective mass varying with $k_z$:
\begin{equation} 
\varepsilon_{\mathbf{k}}^{\pm}=\frac{1}{2}\gamma_2\Gamma^2\pm\frac{k_{\rho}^2}{2m^*(k_z)},
\label{simspec2}
\end{equation}
where 
\beq
m^*(k_z)=\frac{\Gamma \gamma_1}{2v_{\rho}^2}.\label{mass}\eeq
The Fermi energy, $\varepsilon_F$, is determined by a balance of the two terms in Eq.~(\ref{simspec2}), each of which is of order $\varepsilon_F$. Therefore, we did not neglect $\gamma_2$ in the first term, as this would have rendered $\varepsilon_F$ to zero, but neglected it in the second term, where it it would have only given a 
small correction to the effective mass.

Following the tradition, we will refer to the $\pm$ branches of the spectrum as electron/hole bands, although in fact the bands have a mixed electron-hole character, depending on the magntitude and direction of ${\bf k}$. For example, the effective mass for in-plane motion is positive (negative) for the $+$($-$) band, hence the $+$ ($-$) band corresponds to in-plane electrons (holes).  On the other hand, the two bands are degenerate at $k_\rho=0$, while the effective mass along the $c$ axis is positive for $|k_z|<\pi/2c$ and negative for $\pi/2c<|k_z|<\pi/c$ (recall that $\gamma_2<0$). 

In the non-relativistic approximation, the electroneutrality condition fixes the Fermi energy to $\varepsilon_F=(4/3)\gamma_2\approx -26$\;meV, very close to the commonly accepted value of $-25$\; meV.\cite{graphite}  Massless Dirac fermions with dispersions $\varepsilon_{\bf k}=-v_\rho k_\rho$, located near the $H (H')$- points, do not modify significantly the charge balance. However, as we will show in the next Section, the vanishing mass of the in-plane motion near the $H (H')$- points affects dramatically both the longitudinal and Hall conductivities.

Before moving on to the calculation of the conductivities, we need to justify the neglecting $\gamma_3$ in Eq.~(\ref{eq:spectrum}). A perturbation theory in $\gamma_3$ results in a trigonally warped spectrum: \cite{mcc,lmr}
\bea
\varepsilon_{\mathbf{k}}^{\pm}&=&\frac{1}{2}\gamma_2\Gamma^2\nonumber\\
&\pm&\left\{ \frac{k_{\rho }^{2}}{2m^*(k_{z})}+\frac{\sqrt{3}}{2}\gamma _{3}ak_{\rho}\Gamma
 \mathrm{cos}(3\phi )+\frac{\gamma _{1}\gamma _{3}^{2}\Gamma ^{3}}
{2\gamma _{0}^{2}}\mathrm{sin}^{2}(3\phi )\right\}\nonumber,  \label{specgamma3}
\eea
where $\phi$ is the azimuthal angle, and all the smaller bandstructure parameters were neglected in the same way as in Eq.~(\ref{simspec2}). Not too close to the $H (H')$- points, i.e., for $|\Gamma|\sim 1$, the trigonal corrections
 are smaller, though not in order of magnitude, than the first term [$k_{\rho }^{2}/2m^*(k_{z})\sim\varepsilon_F$].
 At the K point, for example,  the second term amounts to $\sim \gamma_3/\gamma_0\sqrt{\gamma_1/\varepsilon_F}\varepsilon_F\approx0.45\varepsilon_F$ and the third term to $\sim
(\gamma_3/\gamma_0\sqrt{\gamma_1/\varepsilon_F})^2\varepsilon_F\approx 0.23\varepsilon_F$, so that the inclusion of $\gamma_3$ does lead to non-negligible changes in the spectrum. 
Indeed, trigonal warping has been found to have non-trivial effects on certain physical quantities such as magneto-optical Kerr rotation and weak-localization (in bilayer graphene).\cite{fal,lev,kec} However, as we discuss in the next section, linear magnetoresistance in weak fields, which is the central result of this work, occurs in a regime where the transport is dominated by the carriers near the
$H (H^{\prime})$ points. Since the trigonal corrections are proportional to a power of $\Gamma$,
they become much smaller
 near the $H (H^{\prime})$- points, and, therefore, $\gamma_3$ can indeed be neglected.  It is true that the inclusion of $\gamma_3$ will produce changes in the magnetoresistance at other regimes, where  magnetotransport is not dominated by carriers near the H- points; however, the effect is expected to be quantitative rather than qualitative, i.e., within the the Boltzmann-equation approach, no new non-analyticities can arise due to $\gamma_3$. 
 Quantitative effects of trigonal warping on MR 
 were
 analyzed 
 in the past; see, e.g., Ref.~\onlinecite{ono2}.

\section{\label{sec3} Linear magnetoresistance}

\subsection{\label{subsec3.1} Magnetoconductivity 
for the non-relativistic energy spectrum}

We now calculate
the components of the magnetoconductivity tensor
in the non-relativistic approximation for the energy spectrum described by Eq.~(\ref{simspec2}). 
We use 
the linearized Boltzmann equation 
in the
relaxation time approximation,
\beq
e\mathbf{E}\cdot\mathbf{v}\frac{\partial f^0}{\partial\varepsilon}=\left(\frac{1}{\tau}-e(\mathbf{v}\times \mathbf{B})\cdot\frac{\partial}{\partial\mathbf{k}}\right)g(\mathbf{k}),
\label{eq:boltz}
\eeq
where $e>0$ is the magnitude of the electron charge, $f^0$ is the Fermi function, and $g({\bf k})$ is the non-equilibrium part of the distribution function.
In general, the Boltzmann equation cannot be solved in a closed form for an arbitrary spectrum (even within the relaxation-time approximation). However, since Eq.~(\ref{simspec2}) is isotropic in the in-plane direction, one can solve Eq.~(\ref{eq:boltz}) exactly for the case of the magnetic field along the normal to the plane. 
Assume that $g(\mathbf{k})=\mathbf{v}\cdot\mathbf{A}$, where $\mathbf{A}$ is an 
in-plane vector represented  in terms of ${\bf E}$ and ${\bf B}\times{\bf E}$ as ${\bf A}=a{\bf E}+b{\bf B}\times{\bf E}$. \cite{zim} 
Then $g({\bf k})$ contains only the in-plane velocity ${\bf v}_{\rho}={\bf k}_{\rho}/m^*(k_z)$. A simplifying feature of the problem with in-plane isotropy is that the coefficients $a$ and $b$  are allowed to be functions only of $k_z$ but not of ${\bf k}_\rho$.  Substituting $g(\mathbf{k})$ into Eq.~(\ref{eq:boltz}) and solving for $a$ and $b$,
we obtain
\begin{equation}
\mathbf{A}=\tau e\frac{\partial f^0}{\partial\varepsilon}
\frac{\m{E}
+\frac{e\tau}{m^*(k_z)}\m{B}\times\m{E}}{1+\left(\frac{e\tau}{m^*(k_z)}\right)^2B^2}.
\label{eq:g}
\end{equation}
The diagonal components of the magnetoconductivity for each of the electron and hole bands are given by
\beq
\sigma^{\pm}_{xx}(B)=\sigma^{\pm}_{yy}(B)=\frac{4e^2\tau}{(2\pi)^3}\int \frac{v_x^2}{1+\omega^2_c(k_z)\tau^2}\left(-\frac{\partial f^0}{\partial\varepsilon}\right)d\m{k},
\label{eq:sigmaxx1}
\eeq
where the integral is over the BZ, the factor of 4
accounts for spin and valley degeneracies, and $\omega_c(k_z)=eB/m^*(k_z)$.
After integrations over ${\bf k}_\rho$ and the azimuthal angle at $T=0$, the previous equation is reduced to
\beq
\sigma_{xx}^{\pm}(B)=\sigma_{xx}^{\pm}(0)\left[1\mp\frac{4e^2\tau}{\pi^2c}\alpha^2\int_{\theta_{\min}}^{\theta_{\max}}d\theta
\frac{(\varepsilon_F-2\gamma_2\cos^2\theta)}{\mathrm{cos}^2\theta+\alpha^2}\right],
\label{eq:sigxxdiv}
\eeq
where $\sigma_{xx}^{\pm}(0)$ is the zero-field conductivity, $\theta=k_zc/2$ and 
\beq
\alpha\equiv\frac{e\tau B}{m^*(0)}\eeq is the dimensionless parameter distinguishing
between the regimes of classically weak ($\alpha\ll 1$) and strong ($\alpha\gg 1$) magnetic fields.
The limits of integration in Eq.~(\ref{eq:sigxxdiv}) are different for the electron and hole bands: 
for electrons, $\theta_{\min}=0$ and $\theta_{\max}=\theta_0$; for holes,  $\theta_{\min}=\theta_0$ and $\theta_{\max}=\pi/2$,
where
$\theta_0=\mathrm{cos}^{-1}\sqrt{\varepsilon_F/2\gamma_2}\approx 0.61$ 
corresponds to $k_z$ at which the Fermi energy intersects the bands (at $k_{\rho}=0$, cf. Fig. ~\ref{fig_fermi}).
Likewise, the off-diagonal components are given by 
\bea
\sigma_{xy}^{\pm}(B)=-\sigma_{yx}^{\pm}(B)&=&\mp \frac{4e^2\tau}{\pi^2c}\alpha\label{xy}\\\
&&\times\int_{\theta_{\min}}^{\theta_{\max}}d\theta
\frac{\cos\theta(\varepsilon_F-2\gamma_2\cos^2\theta)}{\mathrm{cos}^2\theta+\alpha^2}.\nonumber\eea

The origin of the non-analytic dependence of the magnetoconductivity on the magnetic field is evident already from Eqs.~(\ref{eq:sigxxdiv}) and (\ref{xy}). Indeed, if the effective mass 
were
independent of $k_z$, the weak-field behavior of $\sigma_{xx}$ and $\sigma_{xy}$ could 
be
obtained by expanding Eqs.~(\ref{eq:sigxxdiv}) and (\ref{xy}) in $\alpha$. In our case, however, the region of integration over $\theta$ for holes includes the $H (H')$- point ($\theta=\pi/2$), where the effective mass vanishes. An attempt to expand Eqs.~(\ref{eq:sigxxdiv}) and (\ref{xy}) in $\alpha$ leads to the $(\pi/2-\theta)^{-1}$ and $\ln(\pi/2-\theta)$ 
divergences
 in $\sigma_{xx}^-$ and $\sigma_{xy}^{-}$, correspondingly. Cutting off these divergences at $\pi/2-\theta\sim\alpha$, we find that 
 $\Delta\sigma_{xx}^-(B)\equiv \sigma^{-}_{xx}(B)-\sigma^{-}_{xx}(0)\propto |B|$ and $\sigma_{xy}^-(B)\propto B\ln|B|$. For electrons, the corresponding quantities are analytic: $\Delta\sigma_{xx}^+(B)\propto B^2$ and $\sigma_{xy}^+(B)\propto B$. In the strong-field regime, $\sigma_{xx}^{\pm}\propto 1/B^2$ and $\sigma_{xy}^{\pm}\propto 1/B$, as is expected from the Drude model.

Integrals in Eqs.~(\ref{eq:sigxxdiv}) and (\ref{xy}) can be solved for arbitrary $\alpha$. After some algebra, we obtain for the sum of the electron and hole contributions 
 $\sigma_{ij}=\sigma^{+}_{ij}+\sigma_{ij}^-$:
\begin{widetext}
\begin{subequations}
\begin{eqnarray}
\sigma_{xx}(B)&=&\sigma_{xx}(0)\left[1-
\beta\left\{\frac{2|\alpha|}{\sqrt{1+\alpha^2}}\left(1+\frac{3}{2}\alpha^2\right)\left(\frac{\pi}{4}-\tan^{-1}\left[\frac{|\alpha|\mathrm{tan}\theta_0}{\sqrt{1+\alpha^2}}\right]\right)-3\alpha^2\left(\frac{\pi}{4}-\theta_0\right)\right\}\right],
\label{xxf}\\
\sigma_{xy}(B)&=&-\sigma_{xx}(0)\frac{\beta}{2}\alpha\left[\frac{1+\frac{3}{2}\alpha^2}{\sqrt{1+\alpha^2}}\mathrm{ln}\left(\frac{\sqrt{1+\alpha^2}+1}{\sqrt{1+\alpha^2}-1}\right)-3\right],
\label{eq:sigxx2}
\end{eqnarray}
\end{subequations}
where $\sigma_{xx}(0)=\sigma_{xx}^++\sigma^{-}_{xx}(0)=\frac{4e^2\tau|\varepsilon_F|}{\pi^2c\beta}$ and $\beta=8/\left[\pi-4\theta_0+6\mathrm{sin}(2\theta_0)\right]=1.26$. 
For $\alpha\ll 1$, the equations above reduce to
\begin{subequations}
\begin{eqnarray}
\label{eq:linear}
\sigma_{xx}(B)&=&\sigma_{xx}(0)\left[1-\beta\left\{\frac{\pi}{2}|\alpha|-\eta\alpha^2+\mathcal{O}(|\alpha|^3)\right\}\right],\\
\sigma_{xy}(B)&=&-\sigma_{xx}(0)\beta\alpha\left[\ln\frac{2}{|\alpha|}-\frac{3}{2}+\mathcal{O}(\alpha^2)\right],
\end{eqnarray}
\end{subequations}
\end{widetext}
where $\eta=2\left(\tan\theta_0+\frac{3\pi}{4}-3\theta_0\right)\approx 1.92$. The leading terms of these expansions coincide with the estimates obtained by cutting off the integrals in Eqs.~(\ref{eq:sigxxdiv}) and (\ref{xy}), and correspond to non-analytic field dependences of the magnetoresistivity: $\Delta\rho_{xx}(B)\propto |B|$ and $\rho_{xy}(B)\propto B\ln|B|$. Notice that $\rho_{xy}>0$ for $\alpha\ll 1$, which indicates that weak-field magnetotransport is dominated by holes. Although we consider a perfectly compensated case, the equality in the number densities of electrons and holes does not necessarily imply that $\sigma_{xy}$ is equal to zero.
This is so because the weak-filed limit of the Hall conductivity in anisotropic conductors is not directly related to the volumes of the electron and hole Fermi surfaces but is given by an average of certain quantity (equal to $v_y^2/m^*(k_z)$ in our case) over the Fermi surface. 
Although our system is compensated, the hole contribution to $\sigma_{xy}$ exceeds that of electrons. In the strong-field limit, however, the Hall conductivity must achieve a universal limit $\sigma_{xy}=(n_e-n_h)/eB$, \cite{abr} which is equal to zero for compensated metals ($n_e=n_h$). Equation (\ref{eq:sigxx2}) shows that this indeed the case: for $\alpha\gg 1$, the $1/B$ term in $\sigma_{xy}(B)$ is absent and $\sigma_{xy}(B)\approx  -2\sigma_{xx}(0)\beta/15\alpha^3\propto 1/B^3$, which is the expected behavior for compensated semimetals in the case when the $1/B^2$ term is not allowed by lattice symmetry. At the same time, $\sigma_{xx}(B)$ shows the usual behavior in the strong-field limit: $\sigma_{xx}(B)\propto 1/B^2$.

In passing we note that, although nonanalytic MR in graphite arises due to Dirac-like fermions near the $H (H')$- points, the effect occurs {\em not} because the spectrum of these Dirac-like fermions is linear in $k_{\rho}$, but because the mass of normal massive fermions vanishes near the $H (H')$- points 
rendering the fermions Dirac-like. 
 In fact, $\sigma_{xx}$ of  strictly two-dimensional Dirac fermions in doped graphene is expected 
to behave as $B^2$ for $T\ll \varepsilon_F$. (A finite-temperature correction to $\sigma_{xx}$ and the leading term in $\rho_{xx}$ scale as $\sqrt{B}$ but are both exponentially small in this temperature regime. \cite{alekseev})

\begin{figure}[t]
\includegraphics[angle=0,width=0.4\textwidth]{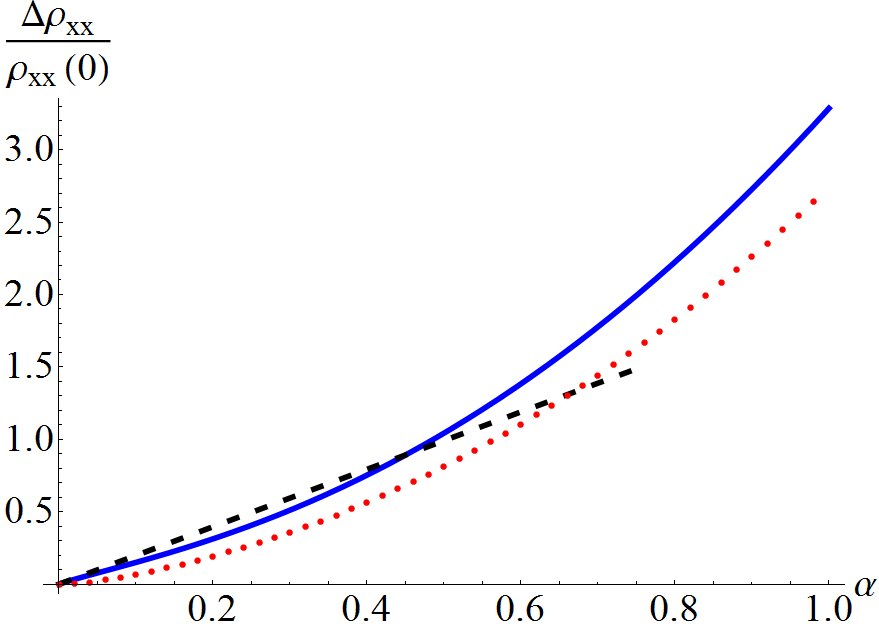}
\caption{
(Color online) Calculated dependence of $\Delta\rho_{xx}/\rho_{xx}(0)=\left[\rho_{xx}(B)-\rho_{xx}(0)\right]/\rho_{xx}(0)$ on the magnetic field in graphite in the weak-field regime. Solid (blue): $\Delta\rho_{xx}(B)$ corresponding to Eqs.~(\ref{xxf}) and (\ref{eq:sigxx2}) for the non-relativistic spectrum,  Eq.~(\ref{simspec2}).  Dotted (red): Numerical results for $\Delta\rho_{xx}(B)$ for the relativistic spectrum, Eq.~(\ref{eq:spectrum1}).
The dashed (black) line shows the asymptotic linear dependence at small fields: $\Delta\rho_{xx}/\rho_{xx}(0)=\beta\pi \alpha/2$. Here $\alpha=e\tau B/m^*(0)$.}
\label{fig1}
\end{figure}

\subsection{\label{subsec3.2} \lq\lq Relativistic\rq\rq\/ effects in the energy spectrum:\\ a new scale for magnetic field}

\begin{figure}[t]
\includegraphics[angle=0,width=0.4\textwidth]{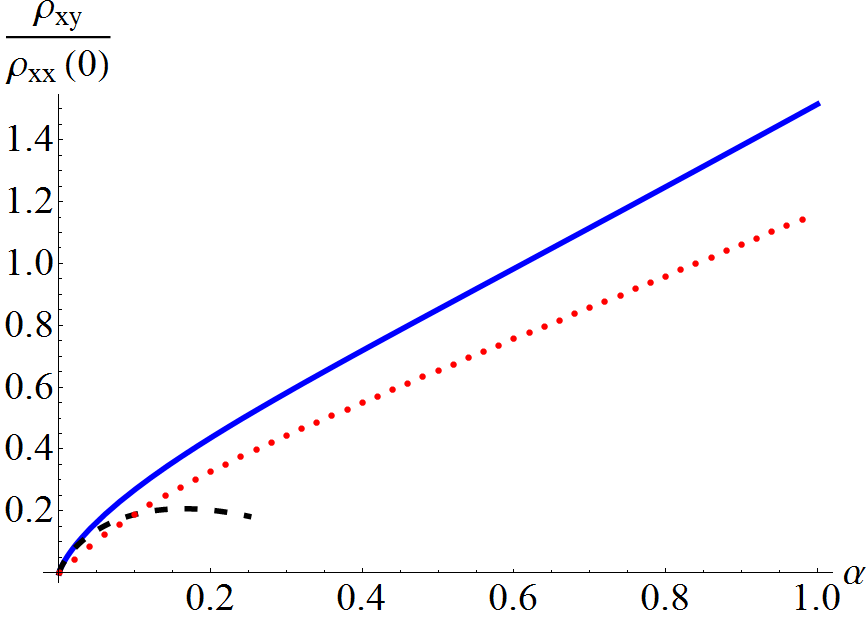}
\caption{
(Color online) Calculated dependence of $\rho_{xy}/\rho_{xx}(0)$ on the magnetic field in graphite in the weak-field regime. Solid (blue): $\rho_{xy}(B)$ corresponding to Eqs.~(\ref{xxf}) and (\ref{eq:sigxx2}) for the non-relativistic spectrum,  Eq.~(\ref{simspec2}).  Dotted (red): Numerical results for $\rho_{xy}(B)$ for the relativistic spectrum, Eq.~(\ref{eq:spectrum1}).
The dashed (black) line shows the asymptotic 
dependence as $\alpha \to 0$: $\rho_{xy}/\rho_{xx}(0)=\beta\alpha\left(\ln\frac{2}{|\alpha|}-\frac{3}{2}\right)$. Here $\alpha=e\tau B/m^*(0)$.}
\label{fig1a}
\end{figure}

In the previous Section, we found that the magnetoconductivity behaves nonanalytically as a function of the magnetic field in the weak-field regime, when transport is controlled by extremely light holes residing near the $H (H')$-points. This result, however, is valid only in the non-relativistic approximation for the energy spectrum, Eq.~(\ref{simspec2}), which is obtained by expanding the relativistic spectrum, Eq.~(\ref{eq:spectrum1}) in $k_\rho$. Returning to Eq.~(\ref{eq:spectrum1}), we see that the actual expansion parameter is the ratio $2v_\rho k_\rho/\gamma_1\Gamma$. The width of the regions contributing to non-analyticities in the magnetoconductivty, $\delta k_z=\pi/c-k_z\sim \alpha/c$, shrinks in proportion to the  magnetic field.   Since $\Gamma\approx c\delta k_z/2
\sim \alpha
\ll 1$ near the $H (H')$- points, we must eventually reach such a weak magnetic field  when $2v_\rho k_\rho/\gamma_1\Gamma\sim 1$, and thus the expansion breaks down. 
Using typical $k_\rho$ estimated as $|\varepsilon_F|/v_\rho$ for Fermi-surface carriers, we find that the expansion is valid only for  
$\alpha_0\equiv |\varepsilon_F|/\gamma_1\ll\alpha\ll 1$. In very weak magnetic fields, such that $\alpha\lesssim|\varepsilon_F|/\gamma_1\approx 0.06$, one needs to consider the complete spectrum in Eq.~(\ref{eq:spectrum1}) and recalculate the dependence of the conductivity on the magnetic field. It is straightforward to show that in this limit the magnetoconductivity is again analytic, i.e., $\Delta\sigma_{xx}(B)\propto B^2
$ and $\sigma_{xy}(B)\propto B$. Since spectrum (\ref{eq:spectrum1}) is isotropic in the $ab$ plane, 
the Boltzmann equation can still be solved analytically. However, the resulting
integrals for the magnetoconductivity need to be solved numerically. Figures \ref{fig1} and \ref{fig1a} show comparisons of the 
magneto-
 and Hall resistivities
 in the relativistic 
 and non-relativistic models. At weak fields, the curves for both the relativistic as well as the non-relativistic models show similar non-analytic dependences. For example, fitting $\Delta\rho_{xx}/\rho_{xx}(0)$ into a linear function of $B$ 
 in the weak-field region
 (we take $0.06\leq\alpha\leq 0.25$  
 since 
 the linearity is most pronounced here) 
 yields  
 slopes that differ by $\approx 30\%$. 
 However, the 
 absolute values of $\Delta\rho_{xx}/\rho_{xx}(0)$ in the two models are different: e.g., at $\alpha=0.2$, the two results differ
 by $\approx 60\%$.
 This difference in the values is due to the fact that at superweak fields  the 
 linear field dependence 
 of the magnetoresistivity predicted by the non-relativistic model is replaced by an analytic, quadratic field dependence predicted by the relaltivistic model thus creating an offset.
Note that the mechanism described above also regularizes a logarithmic
divergence of the Hall constant $R_H=\rho_{xy}/B\propto \ln|B|$ implicit in Eq.~(\ref{eq:sigxx2}).

To summarize, we see that, contrary to the case of conventional metals,  the magnetoconductivity of graphite exhibits three rather than two characteristic regimes: superweak, weak, and strong magnetic fields. 
In the superweak-field regime ($\alpha\ll \alpha_0$), 
the
magnetoconductivity is analytic:
$\Delta\sigma_{xx}(B)\propto B^2$ and $\sigma_{xy}\propto B$. In the weak-field regime ($\alpha_0\ll \alpha\ll 1$), both components of $\sigma$ are non-analytic:
$\Delta\sigma_{xx}(B)\propto |B|$ 
while
 $\sigma_{xy}\propto B\ln|B|$.  
In the strong-field regime ($\alpha\gg 1$), the magnetoconductivity behaves as is expected for a compensated semi-metal. The occurrence of a new scale for the magnetic field, $\alpha_0$, is due to a peculiar feature of the graphite energy spectrum: the existence of Dirac-like fermions near the $H (H')$- points. Notice that 
linear MR,
which is the central result of this section, is still a classically weak-field phenomenon. This should be contrasted to other proposed mechanisms (either classical or quantum), where linear MR occurs
in strong fields.\cite{linmagnature}

\subsection{\label{subsec3.3}Discussion}

\subsubsection{\label{subsec3.3.1} Macroscopic inhomogeneities}

In the previous section, we have shown 
that
the spectrum of graphite allows for linear MR in a certain interval of the magnetic field. 
Using typical values for $\tau$ and band parameters for graphite,
\cite{Du,graphite} we find that the range of linear MR, $\alpha_0\ll \alpha \ll 1$, translates into a range of fields from $\sim 0.006$\;T to $\sim 0.1$\;T. 
In experiment, linear MR is indeed observed in some graphite samples;\cite{kop1,kab2,artunpub} however, it is found to exist not only in this field range but in stronger fields too,  which cannot be accounted for by a simple model of graphite employed in the previous section. Therefore, other factors are also probably at play.

Linear MR in classically strong fields
is often ascribed to macroscopic inhomogeneities in the sample. If the charge carriers in the material, instead of being uniformly distributed, form
macroscopic puddles with different Hall conductivities,
then the effective 
magnetoresistivity
of the sample 
is linear
in classically strong fields.
This idea has been explored in the past,\cite{str1} and recent observations of linear MR in some other materials, such as silver chalcogenides, have been attributed to the presence of such inhomogeneities.\cite{par,str2} It is plausible that linear MR in graphite in strong fields also originates from such inhomogeneities.
A general case of unequal volume fractions has been studied under various approximations \cite{bul,str3}
but, just to illustrate the point, we use the exact result
 \cite{shk}
for the  case of a two-dimensional material 
with equal partial volumes.
(Strictly speaking, the material can still be three-dimensional, only the inhomogeneity has to be two-dimensional). The effective conductivity $\sigma^{e}$ of such a system can be written as
\begin{subequations}
\begin{eqnarray}
\sigma^{e}_{xx}&=&\sqrt{\sigma^{(1)}_{xx}\sigma^{(2)}_{xx}}
\left[ 1+\left(\frac{\sigma^{(1)}_{xy}-\sigma^{(2)}_{xy}}{\sigma^{(1)}_{xx}+\sigma^{(2)}_{xx}}\right) ^2\right] ^{1/2}
\label{eq:sigeffd}
\\
\sigma^{e}_{xy}&=&\frac{\sigma^{(2)}_{xy}\sigma^{(1)}_{xx}+\sigma^{(1)}_{xy}\sigma^{(2)}_{xx}}{\sigma^{(1)}_{xx}+\sigma^{(2)}
_{xx}},\label{eq:sigefft}
\end{eqnarray}
\end{subequations}
where $\sigma^{(i)}$ ($i=1,2$) are the conductivity tensors of the individual components.  

\begin{figure}[t]
\includegraphics[angle=0,width=0.4\textwidth]{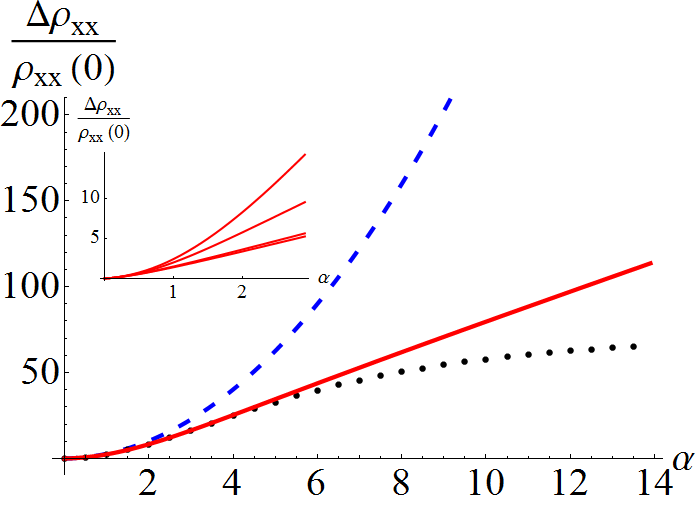}
\caption{
(Color online) Calculated dependence of 
$\Delta\rho_{xx}/\rho_{xx}(0)$
on the magnetic field in graphite over a wide range of fields. 
Solid (red):   
an inhomogeneous system described by Eqs.~(\ref{eq:sigeffd}) and (\ref{eq:sigefft}).
Dashed (blue) and dotted (black): 
compensated and decompensated 
homogeneous systems, respectively.
Here $\alpha=e\tau B/m^*(0)$. \emph{Inset}: Evolution of MR in an inhomogeneous system 
with the degree of decompensation of the two conducting component: $\zeta_i=\delta\varepsilon_{Fi}/\varepsilon_{Fi}$ ($i=1.2)$.
In order of the decreasing slope,  the parameters ($\zeta_1,\zeta_2$) are (0.09, -0.01), (0.2, -0.05), (0.33, -0.33), and (0.4, -0.25).}
\label{fig2}
\end{figure}

It is easy to see from Eqs.~(\ref{eq:sigeffd}) and (\ref{eq:sigefft}) how MR becomes linear in strong fields: if $\sigma^{(i)}_{xx}\propto 1/B^2$ and  
$\sigma^{(i)}_{xy}\propto 1/B
$ in
this regime,
then $\sigma^{e}_{xx}\propto 1/B$ instead of being proportional to $1/B^2$ as in the homogeneous case.  
On the other hand, since $\sigma^{e}_{xy}$ behaves in the usual way, i.e., as 
$1/B$,
 we have $\rho^{e}_{xx}=\frac{\sigma^{e}_{xx}}{\left(\sigma^{e}_{xx}\right)^2+\left(\sigma^{e}_{xy}\right)^2}\propto B$ in strong fields. For graphite, the required $1/B$ dependence of $\sigma^{(i)}_{xy}$ implies that each of the components must be decompensated. Decompensation can be modeled by shifting the Fermi energy away from its value 
for a compensated system. 
We define the degree of decompensation as $\zeta=\delta\varepsilon_F/\varepsilon_F$, 
and choose $\zeta_1=0.09$ and $\zeta_2=-0.01$ for components $1$ and $2$, correspondingly. 
Figure~\ref{fig2} shows $\Delta\rho^{e}_{xx}$ [solid (red)] for an inhomogeneous system. 
The magnetoconductivities of the individual components were calculated using 
the relativistic spectrum from Eq.~(\ref{eq:spectrum1}).
 Also shown is $\Delta\rho_{xx}$ for a compensated [dashed (blue)] and decompensated [dotted (black)] system. 
(We used $\zeta=0.05$ for a homogeneous decompensated system.) 
Comparing the solid and dashed curves, we see how inhomogeneity transforms quadratic MR into linear one.
The dotted line shows that decompensation leads to saturation of MR in a homogeneous system.

It is thus possible that linear
MR
observed in strong fields arises due to extrinsic effects, such as macroscopic inhomogeneities. This effect is distinct from intrinsic linearity discussed 
in Sec.~\ref{subsec3.1},
 which arises due to the presence of Dirac-like holes in graphite and is a weak-field phenomenon. Note that, although intrinsic linearity was shown to exist in perfectly compensated homogeneous graphite, it
is not destroyed
by either decompensation or inhomogeneity--
these two effects
 merely affect the slope 
 of linear MR. Indeed, recall from Eq.~(\ref{eq:linear}) 
 that 
 $\Delta\rho_{xx}(B)/\rho_{xx}(0)\propto \beta\pi\alpha/2$ for weak fields.
  Since $\beta$  is a function of $\varepsilon_F$ (through $\theta_0$), decompensation 
  changes
   $\beta$ and hence the slope. 
   Likewise, 
   inhomogeneities do
   not preclude 
   low-field linear MR but
   modify
     its slope as  $\Delta\rho_{xx}(B)/\rho_{xx}(0)\propto [(\beta^{(1)}+\beta^{(2)})/2][\pi\alpha/2]$, with the superscripts $(1)$ and $(2)$ referring to the two components.
      On the contrary, 
       high-field 
       linear MR 
       occurs only in decompensated {\em and} inhomogeneous graphite.
       Note that the slopes of linear MR are, in general, different in the low- and high-field regimes.
       An important question is whether it is feasible for the two slopes to match within reasonable accuracy, so that MR behaves almost linearly in the entire field range.
    Although one can derive the expressions of analytic results for the two slopes, 
    an analytic comparison of the corresponding expression is quite cumbersome. In lieu of such a comparison, 
        we demonstrate in the inset  of Fig.~\ref{fig2}  how the two slopes 
    evolves for different choices of decompensation in the two components,  As can be seen, for certain decompensations, the slopes in low- and high-field regions are nearly equal. This may explain the variability of the experimental data on linear MR, including those cases where it is observed over a wide range of the magnetic field.

\subsubsection{\label{subsec3.3.2} Quantum magnetooscillations}

Quantum magnetooscillations arise on top of classical MR in strong magnetic fields when $\omega_c$ is larger than not only $1/\tau$ but also temperature. 
Magnetooscillations owe their origin to the extremal cross sections of the
Fermi surface perpendicular to the field. 
In graphite, the extremal (maximal) cross-sections of both the electron and hole Fermi surfaces occur away from the $H (H')$- points. The massive carriers located at these cross-sections give rise to two sets oscillations in the resistivity. However, at the $H (H')$- points where the BZ ends in the $z$-direction, the Fermi surface remains open, leading to 
tiny but non-zero (minimal) cross-sections  (cf. Fig.~\ref{fig_fermi}). Therefore, it is reasonable
to ask if quantum oscillations can also arise due to these minimal cross-sections harboring Dirac fermions. \cite{kop2,pot} We answer this question in the negative: within the accepted band structure, it is \emph{not} possible to have a new set of oscillation arising from the Dirac fermions.
To see this, it suffices to recall that the oscillatory part of the conductivity leading to the Shubnikov-de Haas effect (or of magnetization in case of the de Haas-van Alphen effect) is given by
\beq
\sum_{q\ne 0} \int dk_z I_n(k_z) e^{i2\pi qn(k_z)},
\eeq
where $I_n(k_z)$ is a combination of the single-particle Green's functions depending on the quantity being calculated and $n(k_z)$ is the Landau index as a function of $k_z$. In the limit of a large number of
Landau levels $(n\gg 1)$, one computes the integral via the stationary phase approximation,\cite{abr} in which the main contribution to the result comes from those values of $k_z$ where $n(k_z)$ has an extremum. In
case of graphite, the Landau levels with index number $n$ for the Dirac-like Fermions near the $H (H^{\prime})$- points are described by \cite{graphite}
 \begin{equation}
 n\approx \frac{1}{2ev_{\rho}^2B}\left[\varepsilon_n\left\{\varepsilon_n-\Delta-2\gamma_1\mathrm{cos}\left(\frac{k_zc}{2}\right)\right\}\right].
 \label{eq:landau}
 \end{equation}
It is obvious that this expression does not have a non-zero derivative, and hence an extremum value, at the $H (H')$- points where $k_z=\pm\pi/c$. Therefore, no new quantum oscillations are expected to result from the
carriers near the $H (H')$- points. Note that the sole reason for this negative result is an extra factor of 2 in the denominator within the cosine term in Eq.~(\ref{eq:landau}). This factor of 2 results from
Bernal-stacking of the graphene layers in graphite which makes the periodicity in the $z$-direction to be two lattice planes instead of one. If one were to construct artificial graphite by placing graphene layers directly on
top of each other in the AAAA... fashion, Dirac fermions at the $H (H')$- points would indeed give rise to their own set of quantum oscillations. Note also that our argument means only that Dirac fermions do not give rise to new oscillation {\em frequencies}. The issue of the oscillation {\em phase} is beyond the leading-order semiclassical approximation employed here.

\section{\label{sec4}Conclusions}

In conclusion, we have shown that there exists an interval of the magnetic fields in which in-plane magnetoresistance in graphite (with ${\bf B}$
along the $c$-axis and the current in the $ab$-plane)
scales linearly with the field. Extremely light, Dirac-like carriers located near the $H (H')$- points contribute to such a non-analytic behavior. Linear magnetoresistance
occurs
 for classicaly weak fields, unlike other mechanisms where such a behavior
is found in classically strong or even ultraquantum fields. The Hall resistivity also show a non-analytic, $B\ln{|B|}$ field dependence and is 
dominated by holes. However, 
observed linear magnetoresistance sometimes spans over the entire range-- from weak 
to classically strong fields,
and beyond
-- which cannot be accounted for by 
the mechanism described above
and could be due to macroscopic inhomogeneities in 
real
samples. The light carriers near the $H (H')$- points, however, do not give rise to quantum oscillations of their own, the reason being the Bernal 
stacking of graphene planes in graphite.

\begin{acknowledgements}
This work was supported by NSF-DMR-0908029. We thank V. Yudson for stimulating discussions, and also K. Berke, A. F. Hebard, and S. Tongay for discussing with us their experimental observations on the subject which motivated this work.
\end{acknowledgements}

\end{document}